\documentclass[aps,twocolumn,pra,floatfix,groupedaddress,nofootinbib,notitlepage,showpacs,superscriptaddress]{revtex4-1}
\usepackage{amsmath}
\usepackage{times,bm,bbm,bbold,graphicx,graphics,amssymb,amsmath,amsfonts,dsfont,hyperref,color}
\usepackage{mathrsfs,cancel}
\usepackage[utf8]{inputenc}
\usepackage{graphicx}
\usepackage{soul,xcolor}
\setstcolor{red}
\definecolor{mycolor}{rgb}{0.1, 0.1, 0.7}
\usepackage{dsfont}
\usepackage{url}
\usepackage{soul}
\usepackage[vlined,ruled]{algorithm2e}

\DeclareFontFamily{OT1}{pzc}{}
\DeclareFontShape{OT1}{pzc}{m}{it}%
{<-> s * [1.25] pzcmi7t}{}
\DeclareMathAlphabet{\mathpzc}{OT1}{pzc}%
{m}{it}
\hypersetup{
	plainpages=true,
	breaklinks=true,
	hypertexnames=false,
	pageanchor=true,
	colorlinks=true,
	linkcolor={mycolor},
	citecolor={mycolor},
	urlcolor={mycolor},
	anchorcolor={black}
}
\begin{document}
	\title{Time-dependent atomic magnetometry with a recurrent neural network}
	\author{Maryam Khanahmadi}
	\email{m.khanahmadi@phys.au.dk}
	\affiliation{Department of Physics, Institute for Advanced Studies in Basic Sciences, Zanjan 45137, Iran}
	\affiliation{Department of Physics and Astronomy, University of Aarhus, Ny Munkegade 120, DK 8000 Aarhus C, Denmark }
	\author{Klaus M{\o}lmer}
	\email{moelmer@phys.au.dk}
    \affiliation{Center for Complex Quantum systems, Department of Physics and Astronomy, University of Aarhus, Ny Munkegade 120, DK 8000 Aarhus C, Denmark }
	\begin{abstract}
We propose to employ a recurrent neural network to estimate a fluctuating magnetic field from continuous optical Faraday rotation measurement on an atomic ensemble. We show that an encoder-decoder architecture neural network can process measurement data and learn an accurate map between recorded signals and the time-dependent magnetic field. The performance of this method is comparable to Kalman filters while it is free of the theory assumptions that restrict their application to particular measurements and physical systems.
	\end{abstract}
	\pacs{
			07.55.Ge,05.45.Tp,07.05.Mh.
	}
\date{\today}
\maketitle
\section{Introduction}
\label{sec:int}
Quantum systems can be affected by external perturbations such as electric or magnetic fields contributing to the system Hamiltonian, or thermal baths that contribute damping and decoherence processes.
In quantum metrology, the quantum system is probed in order to estimate these perturbations, and a host of results have been obtained about how one can best extract the relevant information from the measurement outcomes \cite{Maccone,Maccone2}. Special attention is currently devoted to the use of non-classical states, including squeezing and entanglement, and quantum many-body states using, e.g., their critical behavior around phase transition \cite{Caves,Zanardi,Caves2,Hosten,Rams,pezze}. If the task is to estimate a time-dependent perturbation, one should apply continuous or repeated measurements on the quantum probe, and thus follow the dynamical evolution of its quantum state conditioned on the time series of random measurement outcomes. In the absence of unknown perturbations, this dynamics is described by the theory of quantum trajectories \cite{2-optic}. Before their introduction in quantum optics the pertaining stochastic master equations were derived by Belavkin and referred to as filter equations \cite{Belavkin}, and Belavkin also formulated hybrid classical-quantum filter equations \cite{Belavkin2}, which at the same time update the quantum state and the probabilistic information about unknown parameters contributing to the system dynamics \cite{1-filter}. In short, these filters employ a combination of Borns rule which provides the probability of measurement outcomes conditioned on the candidate quantum state, and Bayes rule for the update of prior probabilities for candidate parameter values, conditioned on the given outcome, $p_{\mathrm{Bayes}}(\mathrm{param}|\mathrm{outcome}) \propto p_{\mathrm{Born}}(\mathrm{outcome}|\mathrm{param})\times p_{\mathrm{prior}}(\mathrm{param})$. For many experiments, however, filter theory may not be a realistic means of analysis because its modeling of the system dynamics and measurement statistics and correlations by stochastic master equations may not be valid or practically feasible, for inclusion of complicating features such as finite detector bandwidths, saturation and dead time, and added noise, see, e.g., \cite{Wisman}.

In this paper we study an alternative approach, namely the use of machine learning (ML) to estimate time-dependent perturbations from measurement data. ML techniques provide powerful tools with remarkable ability to recognize and characterize complicated, noisy, and correlated data. ML has made impressive achievements in numerous difficult tasks, such as data mining, classification, and pattern recognition \cite{3-ML}. Neural networks (NNs), as one of the modern tools of ML, have enabled great progress in modeling complicated tasks such as language translation, image and speech recognition \cite{10,12,13-translation,14-speech}. The latter tasks have several similarities with the estimation of a time-dependent perturbation from time-dependent measurement data; this motivates the present study. Note that the NN does not employ any knowledge about the properties of the probe system, how it interacts with the perturbation, or how the data is obtained and processed by the laboratory hardware. We merely assume that we can expose the NN to large amounts of training data obtained under equivalent physical conditions. As an example, we provide measurement data obtained by quantum trajectory simulation of the probing of an atomic ensemble subject to magnetic field that fluctuates in time according to an Ornstein-Uhlenbeck process\cite{Gardiner}.

NNs and their combinations in the so-called deep neural network (DNN) \cite{18-deep} have been used in quantum physics to model quantum dynamics \cite{1-seddiqi,1-carleo,1-banchi,Vicentini,Yoshioka,Nagy}, phase and parameter estimation \cite{1-Carrasquilla,1-klaus,Nolan,1-liu}, and quantum tomography \cite{1-tomography,1-troyer}, and they are increasingly employed in calibration and feedback tasks in quantum experiments \cite{S1,S2,S3,S4}. To our knowledge, the present study is the first to use a deep neural network for the estimation of a stochastically varying perturbation such as a magnetic field interacting with a continuously monitored atomic system.

The paper is structured as follows : In Sec. \ref{background} we describe the atomic system and our procedure to simulate the fluctuating perturbation and a realistic measurement signal. In Sec. \ref{RNN} we describe the structure and motivate our choice of DNN to analyze time-dependent experimental data. In Sec. \ref{result} we present quantitative results and discuss the performance of the DNN for estimation of the magnetic field. In Sec. \ref{conclusion} we conclude and present an outlook.

\begin{figure*}[ht!]
	\includegraphics[scale=.32]{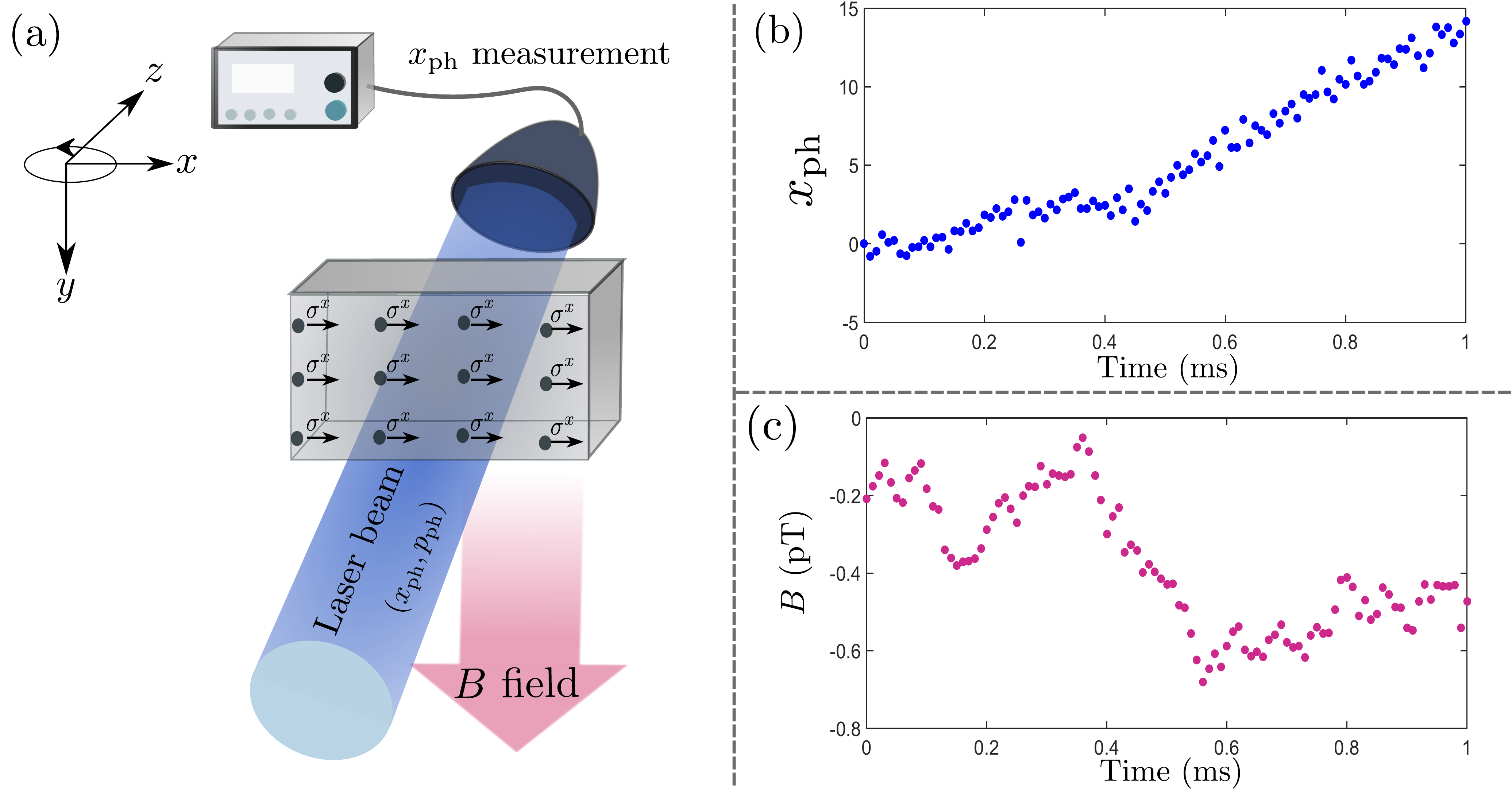}
	\caption{(a) Schematic of the physical model. The atomic spins are polarized along the $x$ direction and interact with a stochastic magnetic field directed along the $y$ axis. The interaction of the system and the magnetic field causes a Larmor rotation towards the $z$ axis. The system is probed by a laser beam propagating along $z$ and the polarization rotation, represented by a canonical field quadrature variable $x_{\mathrm{ph}}$, is measured (see text). Panel (b) shows an example of such a simulated Faraday rotation signal $x_{\mathrm{ph}}$.  Panel (c) shows the noisy magnetic field $B(t)$ that causes the atomic precession and gives rise to the optical signal in panel (b).  }
	\label{f0}
\end{figure*}
\section{Physical model}\label{background}

The experimental scheme is presented in Fig.1(a).
We consider atoms with two Zeeman ground sublevels, equivalent to a spin-$1/2$ particle, which is polarized along the $x$ axis. A large ensemble of such atoms is described by a collective spin operator ($\hbar=1)$, $\vec{J} = (1/2)\sum_{j}\vec{\sigma}^{j} $, where $ \vec{\sigma}^{j} $ is the vector of Pauli matrices describing the individual two-level atoms.

We assume that the atoms are subject to a time-dependent magnetic field directed along the $ y $ direction, which causes a Larmor rotation of the atomic spin toward the $ z $ axis. In our example, the magnetic field fluctuates stochastically as given by an Ornstein-Uhlenbeck process \cite{Gardiner},
\begin{align}
\label{B}
\mathrm{d} B(t)=-\gamma_{b}B(t)\,\mathrm{d}t+\sqrt{\sigma_{b}}\,\mathrm{d}W_{b},
\end{align}
with steady state zero mean  and variance $ \sigma_{b}/2\gamma_{b} $. The noisy Wiener increment $\mathrm{d}W_{b} $ has a Gaussian distribution with mean zero and variance $\mathrm{d}t$. A random realization of Eq. \eqref{B} is shown in Fig. \ref{f0} (c).

For a large number of spin polarized atoms $ N_{\mathrm{at}} $, we may employ the Holstein-Primakoff approximation \cite{Holstein} and introduce effective canonical position and momentum variables as $ x_{\mathrm{at}} = J_{y}/\sqrt{\langle J_{x}\rangle},p_{\mathrm{at}}=J_{z}/\sqrt{\langle J_{x}\rangle} $ , respectively. If the depolarization during the interaction is small, the system retains its spin polarization along the x axis,  $ \langle J_{x} \rangle= \hbar N_{\mathrm{at}}/2 $. The interaction Hamiltonian between the atoms and the magnetic field can then be written $ H_{\mathrm{at}-B}(t)=  \beta B(t) J_{y} =  \mu B(t) x_{\mathrm{at}}$, where $ \beta $ is the magnetic moment, and $\mu \propto \beta \sqrt{ N_{\mathrm{at}}} $.

To probe the system, we consider an off-resonant laser beam which propagates along the $ z $ axis and is linearly polarized along $ x $ direction. This field can be decomposed as a superposition of orthogonal $\sigma^+$ and $\sigma^-$ circularly polarized light, which interact dispersively with the atoms and acquire different phase shifts depending on the state occupied by the atoms. A difference in the phase shift accumulated by the two field components during their passage of the atomic ensemble translates into a (Faraday) rotation of the optical polarization. This rotation is proportional to the population difference between the two atomic sublevels, and hence the collective $J_{z}$ component which, in turn, depends on the magnetic field.

Since both the optical fields and the atomic collective spin are quantum degrees of freedom, the measurements are subject to randomness, and the systems are subject to measurement back action. These are important components that govern the dynamics and the measurement signal and thus the ability to estimate the time-dependent magnetic field. A quantum description of this system by means of Gaussian Wigner functions fully characterized by their mean values and covariances is presented in \cite{11,Mabuchi} and, with special attention to a time dependent $B(t)$ in \cite{4,Cheng}. 

The atom-light interaction and the optical detection occur continuously in time. We discretize this continuous interaction in small time intervals $\tau $.
It is convenient to represent any short segment of the optical field by the Stokes vector of operators $ \vec{S} $ with the $x$ component, $\langle S_{x} \rangle=  N_{\mathrm{ph}}/2$, where $ N_{\mathrm{ph}} $ is the number of photons (during a time interval $ \tau $). During each short-time interval, we consider canonical field variables $ x_{\mathrm{ph}}=S_{y}/\sqrt{\langle S_{x}\rangle},p_{\mathrm{ph}}=S_{z}/\sqrt{\langle S_{x}\rangle} $, and write the interaction Hamiltonian of the atom and light as $ H_{\mathrm{at-ph}}= \kappa p_{\mathrm{at}}p_{\mathrm{ph}}/ \sqrt{\tau} $ where $\kappa \propto \sqrt{N_{\mathrm{at}}\Phi}$ is the atom-light coupling strength
($\Phi=N_{\mathrm{ph}}/\tau$ is the photon flux). This yields the effective total Hamiltonian
\begin{align}\label{e1}
H(t) =  \kappa p_{\mathrm{at}}p_{\mathrm{ph}}/\sqrt{\tau}+\mu B(t)x_{\mathrm{at}}.
\end{align}
During a short time interval $\tau$, the atomic system and field evolve according to Eq. (\ref{e1}) and after each $\tau$, the field observable $ x_{\mathrm{ph}} $ is measured; see Fig. \ref{f0} (b). The subsequent segment of the probing beam, for the next time period $\tau$, is treated by a new set of observables $ (x_{\mathrm{ph}} ,p_{\mathrm{ph}}) $, and the succession of interactions and  measurements thus constitute the continuous measurement scheme. In our analysis below, we shall employ signals obtained by simulation of the magnetic field according to Eq. \eqref{B}, and of the optical measurement according to the distribution of $ x_{\mathrm{ph}} $ due to Eq. \eqref{e1} (i.e., a Gaussian with mean value $ \kappa\sqrt{\tau}\langle p_{\mathrm{at}}\rangle$ and variance $1/2$).

\section{Estimation by a Deep Neural Network}\label{RNN}

The NN approach is ignorant of the atomic system and quantum measurement theory and has the sole purpose to learn from a large collections of simulated realizations of the true magnetic field $ \lbrace \mathbf{B} \rbrace  $ and the corresponding optical detection signals $ \lbrace \mathbf{X}\rbrace  $ at discrete times separated by $ \tau $, how well an unknown time-dependent magnetic field $\mathbf{B} = \lbrace B_{0},B_{\tau},\ldots,B_{t},\ldots,B_{T} \rbrace$ can be estimated from a new signal record $ \mathbf{X} = \lbrace x_{0}, x_{\tau},\ldots,x_{t},\ldots,x_{T}\rbrace$. To this end, we employ the method described in the previous section to prepare a collection of simulated magnetic fields and accompanying noisy optical signals. Both are represented as row vectors with dimension $N_T$ in the total time  $ T $ for each record where $(N_T-1)\tau = T$.

\begin{figure}[ht!]
	\includegraphics[scale=.24]{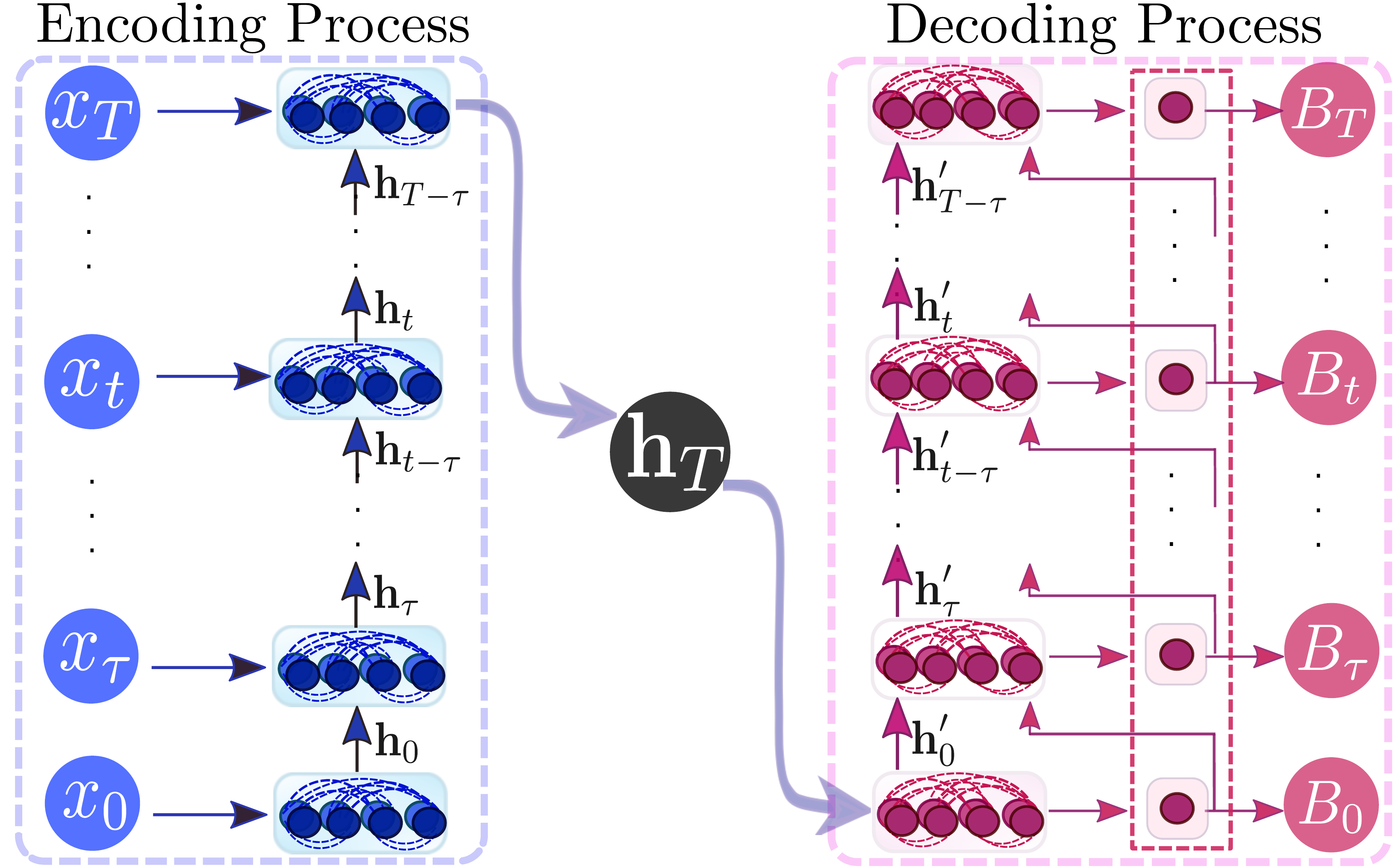}
	\caption{Schematic of the encoder-decoder RNN to process the sequence to sequence translation (measurement data to magnetic field estimate). First, each sequence of the recorded signal $\lbrace x_{0}, x_{\tau},\ldots,x_{T}\rbrace $, is fed to the encoder. The encoder sequentially updates its hidden state $\mathbf{h}_t$, such that the final state holds information about the entire signal from $t=0$ to $t= T$. The final hidden state $\mathbf{h}_T$ is subsequently treated as the initial state of the decoder, which sequentially extracts the predicted magnetic field as its output. Both the encoder and decoder are LSTM cells (see appendix). They have different nonlinear layers which are followed in the decoder by a fully connected layer with one neuron that outputs the predicted value of the field. As shown in the figure, the input of the decoder at each time step $t$ is the previous output estimate $B_{t-\tau}$ and the hidden state $\textbf{h}'_{t-\tau}$.}
	\label{f1}
\end{figure}

The training data $ \lbrace \lbrace \mathbf{X}\rbrace , \lbrace \mathbf{B}\rbrace\rbrace   $, are fed to a  recurrent NN (RNN) which has the ability to extract correlations in sequential data. The characteristic feature of an RNN is its internal (hidden) loop memory allowing it to maintain a state that contains information depending on all previous input as it processes through the sequence of data \cite{9}. This is accomplished by the encoder-decoder architecture shown in Fig. \ref{f1}, which has a remarkable ability to process sequence-to-sequence data \cite{12,13-translation,14-speech} by first processing the input measurement data sequence $ \lbrace \mathbf{X} \rbrace $ to form the hidden vector $ \mathbf{h} $, and to subsequently process $\mathbf{h}$ to form the output candidate estimate of $\lbrace \mathbf{B} \rbrace $. For our purpose, for both the encoder and decoder, we use \textit{Long Short-Term Memory} (LSTM) units which are a specific type of RNNs with ability to process long sequence data \cite{15-LSTM}; for details see the appendix.

Technically, the encoder, at each time step $t$, encodes the information of its input $ x_{t} $ in the internal state $\mathbf{h}_t$ of the NN, which in the LSTM is formally treated by two sets of $m$ nodes $h_{t},c_{t}  $ which are updated according to the nonlinear activation functions

\begin{align}
h_{t} &= \mathpzc{F}(x_{t},h_{t-\tau},c_{t-\tau}),\nonumber\\
c_{t} &= \mathpzc{Q}(x_{t},h_{t-\tau},c_{t-\tau}),
\label{R1}
\end{align}
with a set of weight coefficients $ W_{\mathrm{encoder}} $. After the encoding is finished, the hidden state at the last time step $\mathbf{h}_T = h_T,c_T$ is applied as the initial state $h'_{\mathrm{init}},c'_{\mathrm{init}} = \mathbf{h}_T $ for the decoder. The internal states of the decoder are updated according to previous states and estimated values of the magnetic field as   
\begin{align} \label{R2}
h'_{t} &= \mathpzc{F'}(B_{t-\tau},h'_{t-\tau},c'_{t-\tau}),\nonumber\\
c'_{t} &= \mathpzc{Q'}(B_{t-\tau},h'_{t-\tau},c'_{t-\tau}),
\end{align}
where $ \mathpzc{F'},\mathpzc{Q'} $ are nonlinear functions with weight parameters $W_{\mathrm{decoder}} $; for details of nonlinear functions see appendix. The encoder employs a fully connected layer with one neuron and linear activation function $f(x)=x$. At each time step, a copy of $h'_t$ is fed to the fully connected layer and the candidate subsequent value of the time-dependent magnetic field is obtained with the weight matrix $W''_{h'B} $ and output bias $ b''_{B} $, 
\begin{align} \label{R3}
B_{t} &= f(W''_{h'B}h'_{t}+b''_{B})= W''_{h'B}h'_{t}+b''_{B}.
\end{align}
Note that during the learning procedure, the variable $ B_{t-\tau}$ in Eq. \eqref{R2} is the true magnetic field from training experiments (simulated in our numerical study). But, for the prediction of an unknown signal, it is the estimated magnetic field at the previous step of the encoding process, cf. Eq.\eqref{R3}; see appendix for the details of the  learning and prediction algorithms.

The encoder-decoder RNN is thus fully specified by the elements of the matrices and vectors $W=(W_{\mathrm{encoder}},W_{\mathrm{decoder}},W''_{h'B},b''_{B})$ and it is trained by processing independent sequences of data corresponding to different simulated realizations of the magnetic noise and the optical detection noise. 

The training consists in exposing the RNN to numerous simulated pairs of magnetic fields and measurement data, and minimizing a loss function $ \mathpzc{L} $ which determines the difference between the true $ \lbrace \mathbf{B}_{\mathrm{true}}\rbrace  $ and the predicted field $\lbrace  \mathbf{B}_{ \mathrm{est}}\rbrace$. The loss function is differentiable, and the trainable parameters $ W $ of the RNN are adjusted to minimize $ \mathpzc{L} $ by applying gradient descent steps
\begin{equation}\label{g}
W \leftarrow W - \eta \frac{\partial \mathpzc{L}}{\partial W},
\end{equation}
where $ \eta $ is the learning rate.
According to the learning problem, different kinds of loss function can be applied, e.g., the cross entropy and mean square error which are suitable for classification and regression, respectively. Here, we consider the loss function
\begin{align}
\label{loss}
\mathpzc{L} = \frac{1}{MN_T}\sum_{t=0}^{T}\sum_{i=1}^{M}(B^{i}_{\mathrm{t,true}}-B^{i}_{\mathrm{t,est}})^{2},
\end{align}
where $M$ is the number of the mini-batches on which $ \mathpzc{L} $ is calculated and $N_T$ is the dimension of each sequence. The parameters of the RNN are changed according to the negative direction of the gradient of $ \mathpzc{L} $ until the minimum value is obtained. 

We use the ADAM optimizer \cite{5} to update the parameters of the NN on every mini-batch of size = 256 with the initial learning rate $0.01$. We refer to each iteration over mini batches covering the full data set as an epoch, and we find that applying the gradient ascent for 30 such epochs suffices for convergence of the loss function. We prepare a total of $3\times 10^{6} $ data for which $N = 2.8 \times 10^{6}$ are used as training data and $N' = 2\times 10^{5} $ as unseen (test) data to validate the RNN. Each sequence has $N_T = 101$ dimension from $ t=0$ to $T=1$ with $ \tau = .01 (\,\mathrm{ms}\,)$. As mentioned, we use two LSTMs with the hidden dimension $ m=80 $. The LSTMs alleviate vanishing/diverging gradient problems of simpler RNNs and they improve the learning of long-term dependencies \cite{7-grad}.

\section{Results}\label{result}

\begin{figure*}[ht!]
\includegraphics[scale=.42]{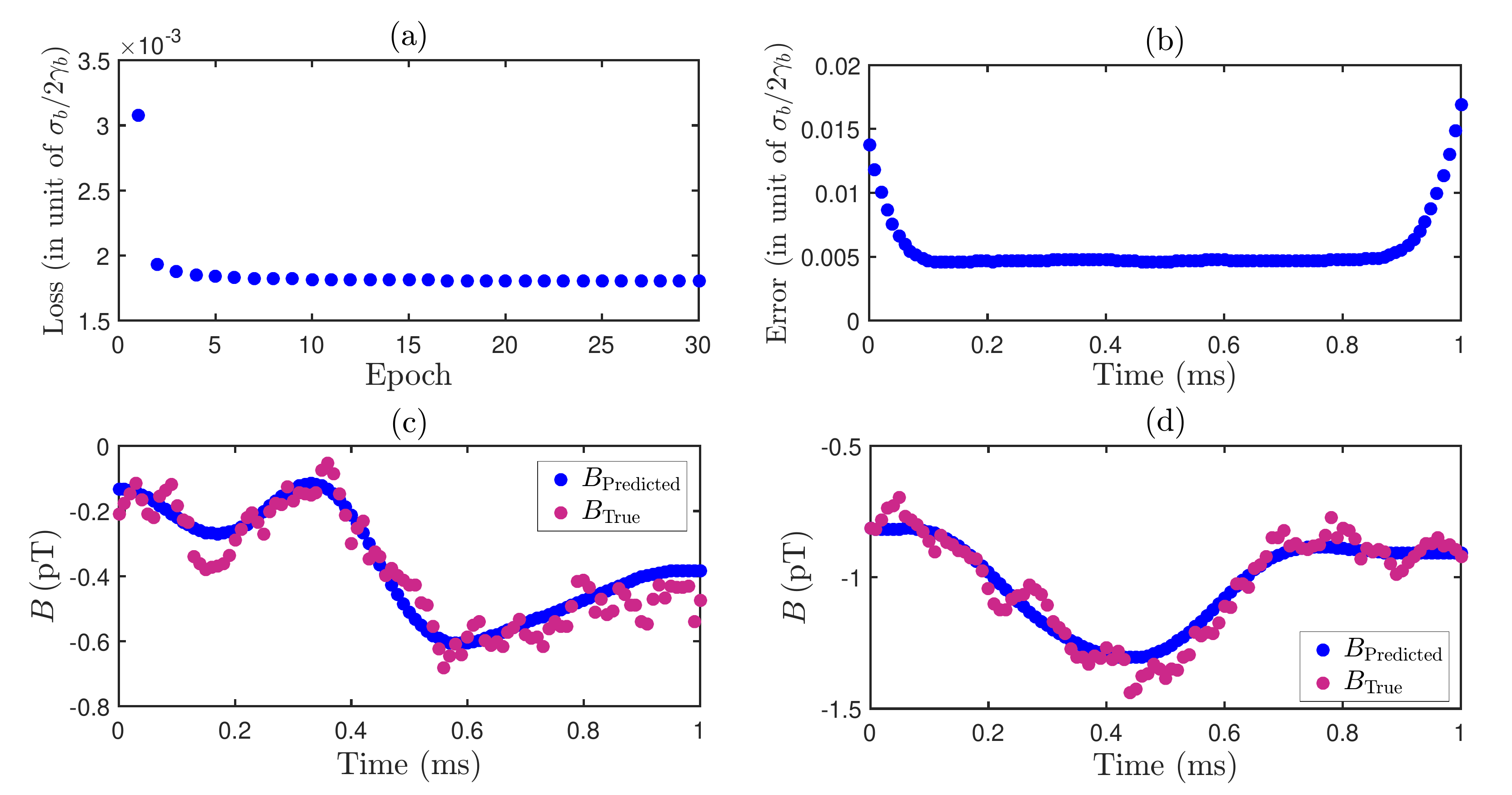}
\caption{(a) The loss function \eqref{loss} is reduced according to the number of epochs of the gradient descent procedure \eqref{g} applied on the training data.  After 30 epochs, the loss function has converged, i.e., it changes by less than or the order of $10^{-6}$ in the last two epochs. (b) The time-dependent mean squared error  \eqref{er} between the predicted and the true magnetic field attains a constant low value inside the probing interval, while increasing at both ends of the time interval, where the lack of prior and posterior measurement data, respectively, causes larger uncertainty. Panels (c,d) show the true magnetic field with red dots together with the value inferred  by the RNN, which is shown by the blue curve. The optical signal leading to the inferred magnetic field $B(t)$ in panel (c) is shown in Fig. \ref{f0}(b). The simulations are made with the following physical parameters, $ \kappa^{2} = 18(\mathrm{ms})^{-1},\mu =90 (\mathrm{ms})^{-1}, N_{\mathrm{at}} = 2\times 10^{12}$ and $N_{\mathrm{ph}} = 5\times 10^{9}$ during each time interval $\tau=0.01$ ms. The magnetic field is sampled with $ \sigma_{b}=2 (pT^{2}/\mathrm{ms}) $ and  $\gamma_{b} = 1(\mathrm{ms})^{-1}$.}
\label{f2}
\end{figure*}
 During the learning procedure, the RNN minimizes the loss function \eqref{loss}, and the learning is iterated until the loss function converges; see Fig. \ref{f2}(a). After the learning process, we validate the NN on the unexplored test data, employing the error function
\begin{align}
\label{er}
\mathrm{Error}(t)=\frac{1}{N'}\sum_{i}^{N'}(B^{i}_{t,\mathrm{true}}-B^{i}_{t,\mathrm{est}})^{2}
\end{align}
where $N'$ is the number of sequences of test data. Eq. \eqref{er} yields the mean squared error of the prediction by the encoder-decoder for the magnetic field from $t = 0$ to $t = T$. The performance of the RNN is shown in Fig. \ref{f2}(b), where the mean squared error is given in units of the variance of the unobserved magnetic field fluctuations.

Despite the similarity of Eqs.\eqref{loss} and \eqref{er}, the loss function shown in Fig. \ref{f2}(a) is smaller than the time average of the error shown in  Fig.\ref{f2}(b). The learning procedure provides the true value of the magnetic field at each time step to the decoder and the loss function concerns only the difference between the true and the guessed value at the next time step. In contrast, after training, the network has access only to the measurement data and thus it applies the previous guessed values at each time step; for the details of the algorithms see \ref{AL}.

Two examples comparing the true and the inferred magnetic field are shown in Fig. \ref{f2}(c), based on the measurement sequence of Fig. \ref{f0}(b), and in Fig. \ref{f2}(d).
The same estimation problem can be treated by Bayesian quantum filter approaches \cite{11,Cheng,Tsang,Tsang2}, and the NN and filter methods yield very similar results. A noticeable feature in Fig. \ref{f2}(b) is the low, almost constant estimation error well within the probing interval, rising to 3-4 times higher values at the ends of the interval. This result is also found in the filter analyses where it is directly associated with the effect of smoothing, i.e., the ability to use both earlier and later measurement data to infer unknown values at any given time.  Earlier and later measurement outcomes are not available at the beginning and the end of the probing time interval. For a simple Gaussian process, this would typically change the variance by a factor of 2, but in our case, both the atomic ensemble state and the magnetic field are unknown, and their mutual correlation is the cause of the larger gain in information by the combination of earlier and later measurements.\\

\section{Summary and conclusions}\label{conclusion}
We have demonstrated the use of a recurrent neural network (RNN) with an encoder-decoder architecture to estimate the value of  a stochastic magnetic field $\mathbf{B}(t)$ interacting with an atomic system during a finite time interval. The input data is the optical signal from continuous Faraday polarization monitoring of the system, which was simulated by quantum trajectory theory. Our RNN learned to estimate  $ \mathbf{B}(t) $ from the optical signal alone and without any information about the dynamics of the system and magnetic field. The performance of our approach is quantitatively similar to the one by a Bayesian analysis of the same data. Although a Bayesian quantum filter approach can also be applied to the example presented here, such an approach relies crucially on several assumptions. Non-Markovian effects related to finite detector bandwidth, saturation and dead time for example  invalidate, or substantially complicate the Bayesian analysis \cite{Wisman}. Other noise processes would not be compatible with a Gaussian distribution of the magnetic field and hence invalidate the strongly simplifying Gaussian description employed in the Bayesian filters \cite{Cheng,4,Tsang}, while they would not impose any fundamental problem for the RNN. Our work confirms the excellent prospects for use of RNN in quantum metrology, and since the goal is always to infer classical functions from classical data, we emphasize that no particular quantum modification of the RNN concept is needed. It is likely that neural networks may be combined with other effective methods, e.g., making use of sparsity in the signal in compressed sensing \cite{IEEE},  as our analysis does not rely on any assumptions about the way the measurement data is provided.
\begin{acknowledgements}
	This work is supported by the Ministry of Science, Research and Technology of Iran, the Danish National Research Foundation and the European QuantERA project C$'$MON-QSENS!.
	The authors acknowledge C. Zhang for sharing her results on quantum filter estimation and A. T. Rezakhani for careful reading of the paper and helpful comments.
\end{acknowledgements}

\appendix
\section*{Appendix: Long short-term memory (LSTM)}
\label{LSTM}
For our RNN we employ the LSTM unit, which is a variant of the simple RNN layer, permitting to process complicated data and carry information across many time steps. The LSTM employs two internal states $ h_{t},c_{t} $, known as the hidden and the cell state respectively, and it processes information by application of three gates, a Forget, an Input/Update and an Output gate.\\

For the encoder (or decoder), at time step $ t $, the input $r_t = x_{t}\,(\mathrm{or}\, B_t) $ and the two internal states $ h_{t-\tau},c_{t-\tau} $ that were updated in previous steps, are fed to LSTM cell. These states and input are employed and updated as follows: The Forget gate $ F $ decides what information to discard from the cell, the Input/Update gate $ I $  decides which input values are employed to update the memory state, and based on the input and the memory of the cell, the Output gate $ O $ decides how the output value is determined. The equations of the gates yielding the subsequent hidden state values $ h_{t},c_{t} $ are  \cite{9,15-LSTM}:
\begin{align}
	I_{t} =& \sigma (W_{ri}r_{t}+W_{hi}h_{t-\tau}+b_{i})\label{l1}\\
	F_{t} =& \sigma (W_{rf}r_{t}+W_{hf}h_{t-\tau}+b_{f})\label{l2}\\
	c_{t} =& F_{t} c_{t-\tau}+I_{t} \tanh(W_{rc}r_{t}+W_{hc}h_{t-\tau}+b_{c})\label{l3}\\
	O_{t} =& \sigma(W_{ro}r_{t}+W_{ho}h_{t-\tau}+b_{o})\label{l4}\\
	h_{t} =& O_{t} \tanh (c_{t}) \label{l5}
\end{align}
where $ \sigma (x)= 1/(1+e^{-x})$ is the Sigmoid function (note that $ \sigma (x)$and $\tanh(x)$ are applied element-wise to each of their arguments). 

\begin{figure}[ht!]
	\includegraphics[scale=.3]{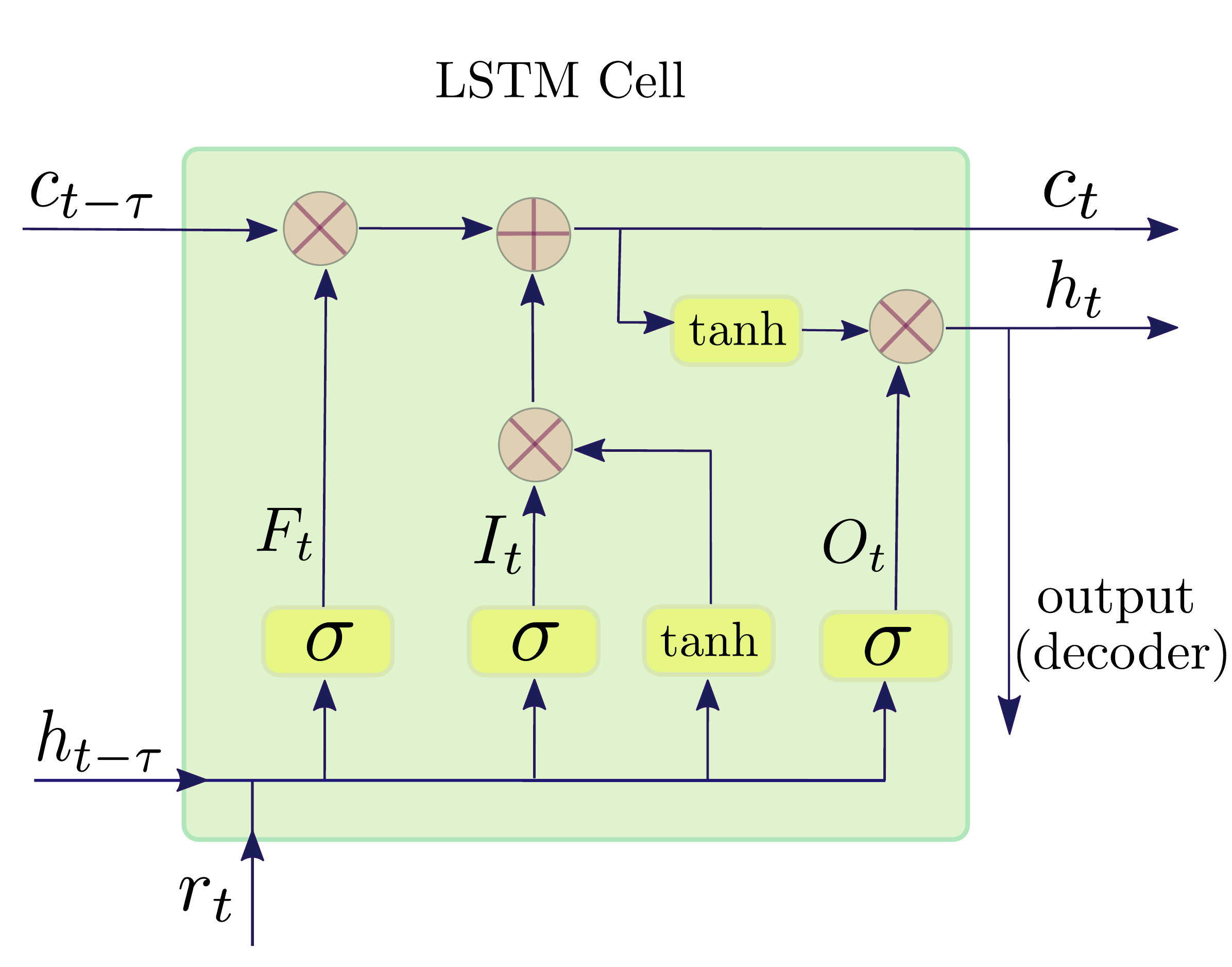}
	\caption{\textbf{Schematic diagram of the LSTM cell.} Each LSTM cell has three inputs where $c_{t-\tau}, h_{t-\tau} $ represent the hidden and cell state at the previous time step $  t-\tau $ and $ r_{t} $ is the input to the cell. Rectangles represent LSTM layers and circles are entry-wise operations; see Eqs. (\ref{l1}--\ref{l5}). For the encoder, the input is the recorded signals and we retain the final hidden and cell state $h_{T}, c_{T}$ as the initial internal state for the decoder by which the output, magnetic field, is predicted. Bifurcations point out the copy operation.}
	\label{f3}
\end{figure}
In the decoder, the LSTM cells are followed by a fully connected layer. As show in Fig. \ref{f3}, one copy of $h_t$ is consider as the LSTM output and is fed to the fully connected layer by Eq. \eqref{R3} whose output is the estimation of the time-dependent magnetic field \eqref{R3}, see Fig. \eqref{f1}. We use LSTMs cells for both the encoder and decoder architecture with hidden dimension $m= 80 $, yielding an $(m\times m)$ matrix $W_{hj}$, an $(m\times 1)$ matrix $W_{rj}$ for $j=i,f,c,o$, and
$m$-dimensional bias vectors. All the parameters $ W,b $ of Eqs. (\ref{l1}--\ref{l5}) are updated in the gradient descent learning of the NN Eq. (\ref{g}) and (\ref{loss}).
The network was implemented by the Keras 2.3.1 and Tensorflow 2.1.0 framework on Python  3.6.3 \cite{16-Keras,17-tensorflow}. All weights were initialized with Keras default.
\section*{Learning and prediction algorithm}\label{AL}
According to Fig. \ref{f1}, two LSTMs are employed to process the training data in two parts: encoding and decoding. In this section we describe the algorithms of the RNN (encoder and decoder) for the learning and prediction procedure.

\begin{algorithm}
\SetAlgoLined
\bigskip
\SetInd{.5em}{1.5em}
 initialize the parameters $ W =(W_{\mathrm{encoder}},W_{\mathrm{decoder}},W''_{h'B},b''_{B})$ according to the Keras default\;
 \For{counter $ < $ epochs}{
 \For {number $ < $ N (training data)}{
  \textbf{"Encoding part"}\\
  initialize $h_{\mathrm{init}},c_{\mathrm{init}},t=0$\;
  \If{$t\leq T$}{
   use $x_t$ from training data as the input\;
   use $h_{t-\tau},c_{t-\tau}$ from previous step\;
   do Eqs. (\ref{l1}-\ref{l5})\;
   keep the updated $h_t,c_t $ for the next time step\;
   $t \gets t+\tau$\;
   }
   keep $h_T,c_T$ as the encoded vector\;
   \textbf{"Decoding part"}\\
   initialize $h'_{\mathrm{init}}=h_T,c'_{\mathrm{init}}=c_T,t=0$\;
   feed an arbitrary initial input $B =0$ to the decoder\;
   \If{$t\leq T$}{
   use $B_t$ from training data as the input\;
   use $h_{t-\tau},c_{t-\tau}$ from previous step\;
   do Eqs. (\ref{l1}-\ref{l5})\;
   keep the updated $h'_t,c'_t $ for the next time step\;
   feed the hidden state $h'_t$ to the fully connected layer\;
   calculate the output $B_t$ of the fully connected layer as Eq. \eqref{R3}\;
   find the square error of the output and the true magnetic field\;
   $t \gets t+\tau$\;
   }
   \If{number is multiplication of mini-batch}{
   evaluate the loss function \eqref{loss} according to square errors calculated in the decoding part\;
   Update parameter $W$ according to Eq. \eqref{g} with ADAM optimizer\;
   }
   }
 }
 \caption{Algorithm of the learning procedure}
\end{algorithm}

\begin{algorithm}[H]
\SetAlgoLined
\bigskip
\SetInd{.5em}{1.5em}
 the parameters \\
 $ W =(W_{\mathrm{encoder}},W_{\mathrm{decoder}},W''_{h'B},b''_{B})$ are learned according to the algorithm 1\;
 \For {number $ < $ N$'$ (test data)}{
  	\textbf{"Encoding part"}\\
  	initialize $h_{\mathrm{init}},c_{\mathrm{init}},t=0$\;
  	\If{$t\leq T$}{
   		use $x_t$ from test (unseen) data as the input\;
   		use $h_{t-\tau},c_{t-\tau}$ from previous step\;
	   do Eqs. (\ref{l1}-\ref{l5})\;
	   keep the updated $h_t,c_t $ for the next time step\;
	   $t \gets t+\tau$\;
   }
   keep $h_T,c_T$ as the encoded vector\;
   \textbf{"Decoding part"}\\ 
   initialize $h'_{\mathrm{init}}=h_T,c'_{\mathrm{init}}=c_T,t=0$\;
   feed an arbitrary initial input $B =0$ to the decoder\;
   \If{$t\leq T$}{
   use $B_t$ from previous output as the input\;
   use $h_{t-\tau},c_{t-\tau}$ from previous step\;
   do Eqs. (\ref{l1}-\ref{l5})\;
   keep the updated $h'_t,c'_t $ for the next time step\;
   feed the hidden state $h'_t$ to the fully connected layer\;
   calculate the output $B_t$ of the fully connected layer as Eq. \eqref{R3}\;
    show the output as the estimated magnetic field\;
   find the square error of the output and the true magnetic field\;
   $t \gets t+\tau$\;
   }
   }
   find the average error, cf.,  Eq. \eqref{er} \;
 \caption{Algorithm of the prediction procedure}
\end{algorithm}
\end{document}